# Angular dependent magnetization reversal in exchange biased bilayers under a modified 'effective field model'

Wei Zhang, Qing-feng Zhan, and Kannan M. Krishnan [*]

*Department of Materials Science, University of Washington, Seattle, Washington 98195, USA*

**Abstract**

A modified effective field model was developed to quantitatively interpret the angular dependent magnetization reversal processes in exchange biased Fe/IrMn bilayers. Several kinds of multi-step loops with distinct magnetization reversal routes were observed for the samples measured at various field orientations. Two types of angular dependent switching fields are observed and their transitions are investigated, which are found to be driven by both Fe and IrMn layer thicknesses. Our modified effective field model can nicely describe all the switching field behaviors including the critical effects of the exchange bias induced uniaxial anisotropy on the magnetization reversal processes.

[*] Author to whom correspondence should be addressed. Email: kannanmk@uw.edu

## I. Introduction

Exchange bias (EB) of ferromagnetic (F) / antiferromagnetic (AF) bilayers has been extensively studied in the past decades due to its key role in magnetic recording technologies[1]. One of the intriguing properties of EB is the asymmetric magnetization reversal behavior[2], however, whose origin still remains controversial. Such asymmetric behavior has been correlated, depending on the nature of the bilayers, with different magnetization processes[2-7], and the angle between the external field and the EB direction[8-10]. To probe such properties, the angular dependent magnetic measurement has been widely adopted with various experimental setups such as the vector magneto-optic Kerr effect (MOKE)[11,12], which offers a comprehensive understanding of the magnetic reversal processes by probing both the longitudinal and transverse magnetization components.

The magnetic reversal processes in magnetic films are determined by the magnetic anisotropies[13]. In polycrystalline EB bilayers, a unidirectional anisotropy, $K_{eb}$, established by EB as well as an induced uniaxial anisotropy, $K_u$, are introduced, which play the key role on altering the magnetic switching process. Considering both $K_{eb}$ and $K_u$, the value of EB, $H_{eb}$, and coercivity, $H_c$, can be numerically fitted by the Stoner-Wahlfarth model[3]. In epitaxial EB systems, the intrinsic symmetry of the magnetocrystalline anisotropy needs to be further considered, which results in the multi-step hysteresis loops and a complex angular dependent behavior[14,15]. An effective field model (EFM), taking into account the unidirectional anisotropy field,

$H_X$, and the cubic ferromagnetic anisotropy field, $H_A$, was proposed to quantitatively interpret the complicated angular dependent switching fields[14]. Recent studies also showed that both $H_A$ and $H_X$ are inversely proportional to the ferromagnetic layer thickness and are therefore interface related effects[15].

In this work, we systematically studied the magnetization reversal behaviors of Fe/IrMn EB bilayers. By tuning the IrMn layer thickness, $t_{IrMn}$, the magnetization reversal characteristic in the case of both $H_A > H_X$ and $H_A < H_X$ were obtained. Different types of angular dependent switching fields were observed for each case and were interpreted by the modified effective field model (mEFM) that considers $H_X$, $H_A$ and an EB induced uniaxial anisotropy $H_U$. It should be noted that $H_U$ in continuous films can also be induced by various origins including surface steps[16] and oblique growth[17,18]. By tuning the Fe layer thickness, $t_{Fe}$, we showed that the different angular dependent behaviors are not a sole function of the anisotropy ratio, $H_A/H_X$ as described in the EFM, but also critically depends on the induced $H_U$, especially when $H_A$ and $H_X$ are of comparable magnitudes. The induced $H_U$ can give out a new set of angular dependent behaviors in the phase diagram that summarizes all the possible angular dependent magnetization reversal processes. The critical effect of the $H_U$ on the magnetization reversals, which are well described by mEFM, is particularly important in magnetic thin-film devices that require accurate controls on the magnetic switchings.

**II. Experiment**

Fe/IrMn bilayers were grown on MgO(001) substrates by Ion Beam Sputtering (IBS) with a base pressure below $1\times10^{-8}$ mbar and a deposition rate as low as 1Å/s[19]. The substrates were pre-annealed at 500 °C for 1.5 hour and held at 130 °C for the deposition of Fe and IrMn layers. A permanent magnet with a magnetic field ~300 Oe was positioned along the Fe[100] direction during growth. Two series of samples (A and B) were grown with the structure of MgO/Fe($t_{Fe}$)/IrMn($t_{IrMn}$)/Ta(3nm cap). Series A has a fixed $t_{IrMn}$ = 4.5 nm, and different $t_{Fe}$ = 10 nm ($A_{10}$), 30 nm ($A_{30}$), 50 nm ($A_{50}$), and 70 nm ($A_{70}$); series B has fixed $t_{IrMn}$ = 8 nm, and different $t_{Fe}$ = 10 nm ($B_{10}$), 30 nm ($B_{30}$), 50 nm ($B_{50}$), and 70 nm ($B_{70}$). X-ray diffraction (XRD) with Cu $K\alpha$ radiation is used to examine the crystal structure of the samples. The (002) peaks of Fe and IrMn in the longitudinal $\theta$–$2\theta$ scan (Fig. 1(a)) indicated a good out-of-plane (002) texture of the samples[20].

## III. Results and discussions

Magnetic properties were characterized at room temperature by combined measurements of the longitudinal (∥) and transverse (⊥) MOKE for different field orientation, $\phi$, which is defined as the angle between the external applied field and the field cooling direction, with $\phi$ increasing in the clockwise direction. The EB gives rise to the unidirectional anisotropy, $H_X$, and a collinear uniaxial anisotropy, $H_U$, along the field cooling direction. Both of them are superimposed on the Fe cubic magnetocrystalline anisotropy (Fig. 1(b)), thus inducing various magnetic switching routes between the Fe easy axes.

Different magnetic reversal processes as well as the angular dependent behaviors are found in $A_{10}$ and $B_{10}$. At $|\phi| < 40°$, the magnetic transitions for $A_{10}$ in the descending branch take place at $H_{c1}$ and $H_{c2}$, with $H_{c1} > H_{c2}$ ($10° < |\phi| < 40°$), Fig. 2(b). The transverse MOKE loop reveals that the first step is an intermediate state in which Fe spins are oriented along the [010] axis for $10° < \phi < 40°$, and along the [0-10] axis for $-40° < \phi < -10°$, when the magnetization switches from the [100] to [-100] directions. The magnetic transition displays a single step at $H_{c1}$, with $H_{c1} < H_{c2}$, at $|\phi| < 10°$, Fig. 2(a). For the ascending branch, however, the magnetization switches smoothly from [-100] to [100] direction without resting at any intermediate state, suggesting that the magnetization reversal occurs by $180°$ DW nucleation[15]. The magnetization reversal for $B_{10}$ in the descending branch is the same process as $A_{10}$. However, for the ascending branch, the direct $180°$ magnetic switching cannot be observed but is replaced by a two-step reversal (Fig. 2(f)), i.e., the magnetic transitions take place at $H_{c3}$ and $H_{c4}$, with $H_{c3} < H_{c4}$ ($15° < |\phi| < 40°$). It should be noted that $H_{c3}$ and $H_{c4}$ belong to the same semicircle as $H_{c1}$ and $H_{c2}$, as revealed by the transverse MOKE signals. When $H_{c3} > H_{c4}$ ($|\phi| < 15°$), only a one-step reversal at $H_{c3}$ can be observed (Fig. 2(e)). Detailed discussions about the two successive magnetic transitions can be found elsewhere[21,22]. For the field applied around the Fe hard axes, i.e., $40° < |\phi| < 50°$, hysteresis loops with only one irreversible transition on both branches are observed for both samples (Fig. 2(c) and (g)). Double-shifted loops with two-stage magnetic transitions occurring on both branches ($H_{c5}$, $H_{c6}$ for the descending branch and $H_{c7}$, $H_{c8}$ for the ascending branch) are observed for both samples at $50^o < |\phi| < 130^o$, far away from the bias direction (Fig. 2(d) and (h)). The transverse MOKE loops show the intermediate state for the ascending and descending branches lie on the same axis given by the bias, i.e., the [100] direction, and the switching route is [0-10] → [100] → [010] for decreasing field and the reverse route for increasing field.

The angular dependent switching fields for $A_{10}$ and $B_{10}$ are summarized in Figure 3. The data is presented in two different field regions, (1) $\phi_{\parallel}$, for the field orientation $\phi = \phi_0 + \triangle\phi$ with $\phi_0 = 0^o$ and $-40^o < \triangle\phi < 40^o$, i.e., close to the bias direction; (2) $\phi_{\perp}$, for the field orientation $\phi = \phi_0 + \triangle\phi$ with $\phi_0 = -90^o$ and $-40^o < \triangle\phi < 40^o$, i.e., close to perpendicular to the bias direction. Here, we use the mEFM to quantitatively interpret the angular dependence of the switching fields. The effective field, $H_{eff}$, is considered to come from four different contributions rather than three as it was originally proposed, i.e., the external magnetic field, $H_{ext}$, the exchange field, $H_X$, the EB induced uniaxial anisotropy field, $H_U$, and the cubic anisotropy field, $H_A$, which is aligned with the Fe easy axis. Notably, both $H_A$ and $H_U$ depend on the projection of the ferromagnetic magnetization onto the Fe easy axis.

The switching fields can be derived by comparing the effective fields at the initial and final Fe easy axes involved in the magnetic transition[14,15]. The theoretical switching fields for $90^o$ magnetic transitions are obtained as:

$$H_{c1} = -(H_X + H_A + H_U)/(\cos\triangle\phi + \sin\triangle\phi),$$

$$H_{c2}= - (H_X+H_A-H_U)/(\cos\triangle\phi - \sin\triangle\phi),$$

$$H_{c3} = -(H_X-H_U-H_A)/(\cos\triangle\phi -\sin\triangle\phi),$$

$$H_{c4} = -(H_X+H_U-H_A)/(\cos\triangle\phi + \sin\triangle\phi),$$

$$H_{c5} = - (H_A-H_U-H_X)/(\cos\triangle\phi - \sin\triangle\phi),$$

$$H_{c6} = - (H_A+H_U+H_X)/(\cos\triangle\phi +\sin\triangle\phi),$$

$$H_{c7} = (H_A-H_U-H_X)/(\cos\triangle\phi + \sin\triangle\phi),$$

$$H_{c8} = (H_A+H_U+H_X)/(\cos\triangle\phi - \sin\triangle\phi).$$

For 180º magnetic switching from the [-100] to [100] axes,

$$H_c = (H_A-H_X+H_U)/\cos\triangle\phi.$$

As can be seen in Fig. 2 and 3, the characterizing switching fields of $A_{10}$ satisfy $H_{c1}<0$, $H_{c2}<0$, and $H_c>0$ within $\phi_\parallel$. A relationship of the effective fields $H_A>H_X-H_U$ can be therefore predicted according to the above equations. Similarly, we found the switching fields $H_{c1}<0$, $H_{c2}<0$, $H_{c3}<0$, and $H_{c4}<0$ for of $B_{10}$, indicating $H_X>H_A\pm H_U$. Within $\phi_\perp$, the switching fields for $A_{10}$ satisfy $H_{c5}$, $H_{c6}< 0$ and $H_{c7}$, $H_{c8}> 0$, thus $H_A > H_X + H_U$. However, the switching fields for $B_{10}$ satisfy $H_{c6}$, $H_{c7}< 0$ and $H_{c5}$, $H_{c8}> 0$, which indicates $H_X > H_A - H_U$. Here, we label the angular dependent switching fields for $A_{10}$ as type-$I_0$ (at $\phi_\parallel$) and type-$I_{90}$ (at $\phi_\perp$) behavior (Fig. 2(a-b)), and for $B_{10}$ as type-$II_0$ (at $\phi_\parallel$) and type-$II_{90}$ (at $\phi_\perp$) behavior (Fig. 2(c-d)). Each type of angular dependent behavior has its distinct characterizing switching fields and effective field relationships (Table 1). Notably, in the original EFM which excludes $H_U$, the effective field relationships regresses to two simple ones, i.e., $H_A>H_X$ for Type-$I_{0,90}$ and $H_A<H_X$ for Type-$II_{0,90}$.

Using the above-derived expressions, the angular dependence of the switching fields for $A_{10}$ and $B_{10}$ can be well fitted (Fig. 3). Our fitting gave the anisotropy fields: $H_A$= 55 Oe, $H_X$ = 19 Oe, $H_U$ = 6 Oe for $A_{10}$; and $H_A$= 20 Oe, $H_X$ = 45 Oe, $H_U$ = 4 Oe for $B_{10}$. Using the relation $J_{ex} = H_X M_{Fe} t_{Fe}$ and the magnetization for bulk Fe, $M_{Fe}$ = 1700 emu/cm$^3$, and $t_{Fe}$ = 10nm, the interface energy, $J_{ex}$, between Fe and IrMn is obtained as 0.032 erg/cm$^2$ for $A_{10}$ and 0.077 erg/cm$^2$ for $B_{10}$. The fitting results further confirmed that both $H_A$ and $H_X$ are interfacial related effects, which also points directly to the competing effects of the pinned and rotatable AF spins at the F/AF interface[23-26]. Specifically, for sample $A_{10}$ with $t_{IrMn}$ = 4.5 nm, the AF anisotropy energy is not strong enough to fully establish the bias. Sufficient numbers of the AF spins at the interface would reverse with the F spins due to the strong exchange coupling, and contribute to the rotatable ferromagnetic anisotropy, $H_A$, rather than be pinned and enhance $H_X$. For sample $B_{10}$, the increased $t_{IrMn}$ (= 8nm) results in more AF spins being pinned and less of them being rotatable. Thus an increase in $H_X$, but a decrease in $H_A$ are observed. This observation is analogous to the competing effect of $H_{eb}$ and $H_c$ in polycrystalline EB bilayers[27].

In the original EFM, the Type-$I_0$ (Type-$II_0$) angular dependent behavior is predicted to show up simultaneously with Type-$I_{90}$ (Type-$II_{90}$) according to the effective field relationships. This has found no problems in the reported Fe/MnF$_2$[14] and Fe/MnPd EB[15] systems; as in both cases $H_A$ is significantly larger than $H_X$. Nevertheless, as we have demonstrated in the mEFM, taking into account the $H_U$ discriminates the critical conditions for the type-$I_0$ and type-$I_{90}$ angular dependent behaviors, as well as for type-$II_0$ and type-$II_{90}$. By comparing those critical conditions, we predict theoretically that the observation of type-$II_0$ should simultaneously lead to the observation of type-$II_{90}$, since $H_X > H_A - H_U \subset H_X > (H_A \pm H_U)$; however, the observation of type-$I_0$ does not necessarily lead to the observation of type-$I_{90}$, because $H_A > H_U + H_X \not\subset H_A > H_X - H_U$. In other words, it is possible to observe the combination of both type-$I_0$ and type-$II_{90}$ when the relationship $H_U + H_X > H_A > H_X - H_U$ is satisfied. This critical effect of $H_U$ is particularly significant when $H_A$ and $H_X$ are of comparable magnitude.

To investigate the critical effect of $H_U$, MOKE angular dependent measurements were further performed on all A and B series samples. For the A series, all samples are characterized by the switching fields $H_c > 0$, $H_{c5} < 0$, $H_{c7} > 0$, and display the type-$I_{0,90}$ behavior. mEFM fittings gave out the relationship $H_A > H_X \pm H_U$ for all samples, $A_{10}$-$A_{70}$. For the B series samples, fully biased hysteresis loops are observed for $B_{10}$ and $B_{30}$ when measuring along the bias direction, but not for $B_{50}$ and $B_{70}$. On the other hand, loops measured perpendicular to the bias direction have the same feature of $H_{c5} > 0$, $H_{c6} < 0$, $H_{c7} < 0$ and $H_{c8} > 0$. The angular dependence of switching fields for $B_{30}$ displays the type-$II_{0,90}$ behavior (Fig. 4(a) and (d)). Compared with $B_{10}$, the evolution via increasing $t_{Fe}$ can be recognized by the larger onset $|\triangle \phi|$ for $H_{c4}$, and also the smaller values of $|H_{c4}|$, $|H_{c5}|$ and $|H_{c7}|$. As the Fe layer thickness further increases, the angular dependent behaviors for $B_{50}$ and $B_{70}$ changed from the type-$II_{0,90}$ to the combination of type-$I_0$ (Fig. 4(b) and (c)) and type-$II_{90}$ (Fig. 4(e) and (f)) behavior, with the 180° magnetic transition taking place in the whole range of $\phi_\parallel$ for the ascending branch. We fitted the curves for each sample and the parameters are summarized in Table 2. Consistently, the anisotropy fields satisfy the relationship of $H_X>H_A+H_U$ for $B_{10}$ and $B_{30}$, but $H_U+H_X>H_A>H_X-H_U$ for $B_{50}$ and $B_{70}$, respectively. As can be seen, the induced $H_U$, though quite small, can also significantly alter the magnetization reversal processes. A phase diagram shows the different sets of angular dependent behavior is varied with the ratio $(H_X-H_U)/H_A$ and $(H_X+H_U)/H_A$, as plotted in Fig. 5. In the original EFM, only two sets of angular dependence, i.e., type-$I_{0,90}$ and type-$II_{0,90}$, are predicted to appear on either side of the critical point of $H_X=H_A$. The induced $H_U$ in our mEFM results in the new set of combination (type-$I_0$, type-$II_{90}$) where $H_U + H_X > H_A > H_X - H_U$ is satisfied.

## IV. Conclusion

In summary, the magnetic properties of Fe/IrMn EB bilayers with different Fe and IrMn thicknesses were characterized using vector MOKE. Square loops, asymmetric-shaped loops, and different two-step loops were observed when measuring at various field orientations. Two sets of typical angular dependent switching fields and their transitions are investigated, which are found to be driven by both Fe and IrMn layer thicknesses. Considering the cubic, the uniaxial, and the unidirectional anisotropies, we developed a mEFM that nicely describes the different magnetization reversal processes together with the two types of angular dependent behaviors. The induced uniaxial anisotropy can significantly alter the critical onsets for the different angular dependent switching fields, as demonstrated both experimentally and theoretically. A phase diagram describing the angular dependent behaviors with the relative magnitudes of the effective fields was also plotted.

## ACKNOWLEDGEMENT

This work was supported by DoE/BES under Grant No. ER45987.

**TABLE and FIGURE CAPTIONS:**

| range | Type | Characterizing switching fields | Effective fields relationship (mEFM) | Effective fields relationship (EFM) |
|---|---|---|---|---|
| $\phi_\parallel$ | Type-I$_0$ | $H_{c1}, H_{c2} < 0, H_c > 0$ | $H_A > H_X - H_U$ | $H_A > H_X$ |
| | Type-II$_0$ | $H_{c1}, H_{c2}, H_{c3}, H_{c4} < 0$ | $H_X > H_A \pm H_U$ | $H_X > H_A$ |
| $\phi_\perp$ | Type-I$_{90}$ | $H_{c5}, H_{c6} < 0, H_{c7}, H_{c8} > 0$ | $H_A > H_X + H_U$ | $H_A > H_X$ |
| | Type-II$_{90}$ | $H_{c6}, H_{c7} < 0, H_{c5}, H_{c8} > 0$ | $H_X > H_A - H_U$ | $H_X > H_A$ |

Table. 1 Summary of the characterizing switching fields and the effective fields relationship of the different types of angular dependent behaviors for mEFM and EFM.

| Sample | $H_A$ / Oe | $H_X$ / Oe | $H_U$ / Oe |
|--------|-----------|-----------|-----------|
| $B_{10}$ | 20 | 45 | 4 |
| $B_{30}$ | 9 | 18 | 3 |
| $B_{50}$ | 8.5 | 7.5 | 3 |
| $B_{70}$ | 7 | 6 | 2 |

Table. 2 Fitted anisotropy fields $H_A$, $H_X$, and $H_U$ of $B_{10}$-$B_{70}$.

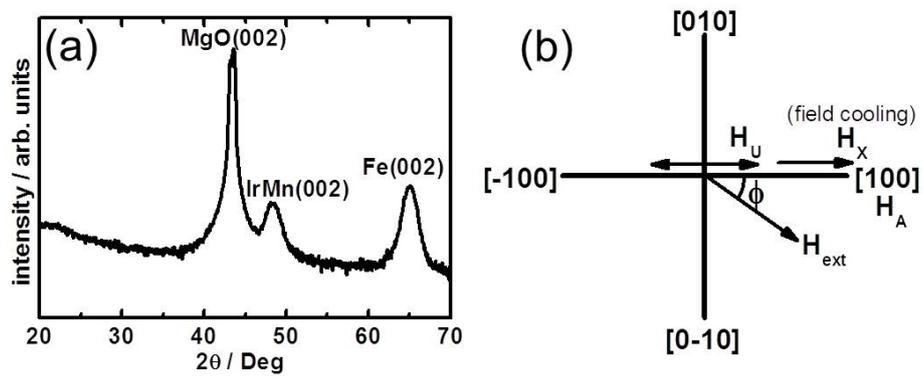

Fig. 1 (a) XRD $\theta$–$2\theta$ scan of Fe(10nm)/IrMn(8nm) EB bilayer. (b) Geometry of the field cooling direction, magnetic anisotropies and the external magnetic field.

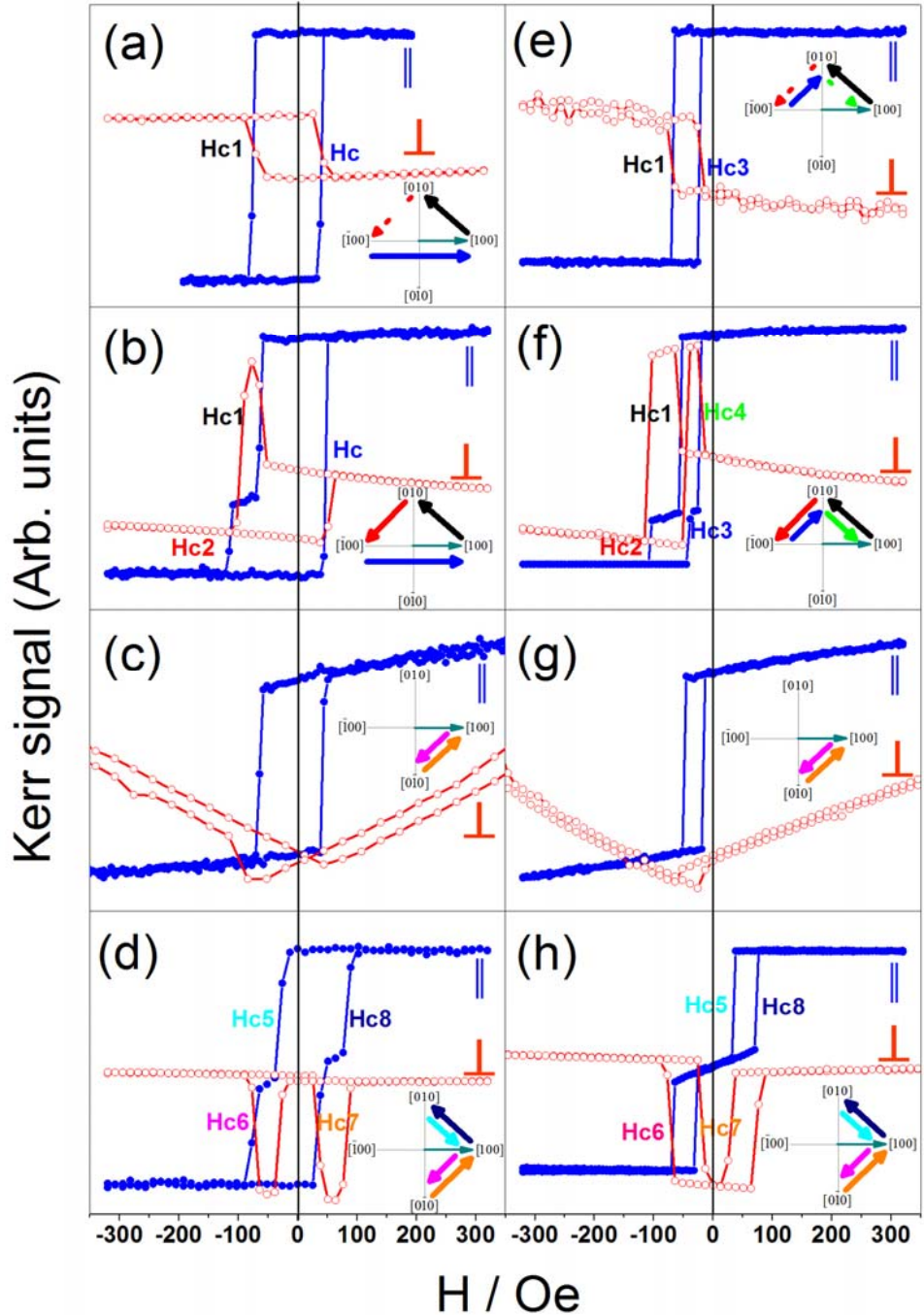

Fig. 2 (color online) Longitudinal (∥) and transverse (⊥) MOKE loops measured at various field orientations at $\phi=2.5°$, $\phi=22.5°$, $\phi=-42.5°$, and $\phi=-87.5°$ for $A_{10}$ (a-d) and $B_{10}$ (e-h). Inset figures illustrate the magnetization reversal process corresponding to the hysteresis loop. The switching fields in each main figure are shown by the arrows with corresponding colors in the inset figure. Dashed arrow indicates the switching field that is not displayed during the two successive magnetic transitions. The asymmetry (hysteresis) in transverse (⊥) signals in (a) and (b), at $\phi=2.5^\circ$, is due to the slight misalignment between MOKE laser beam plane and the magnetic field.

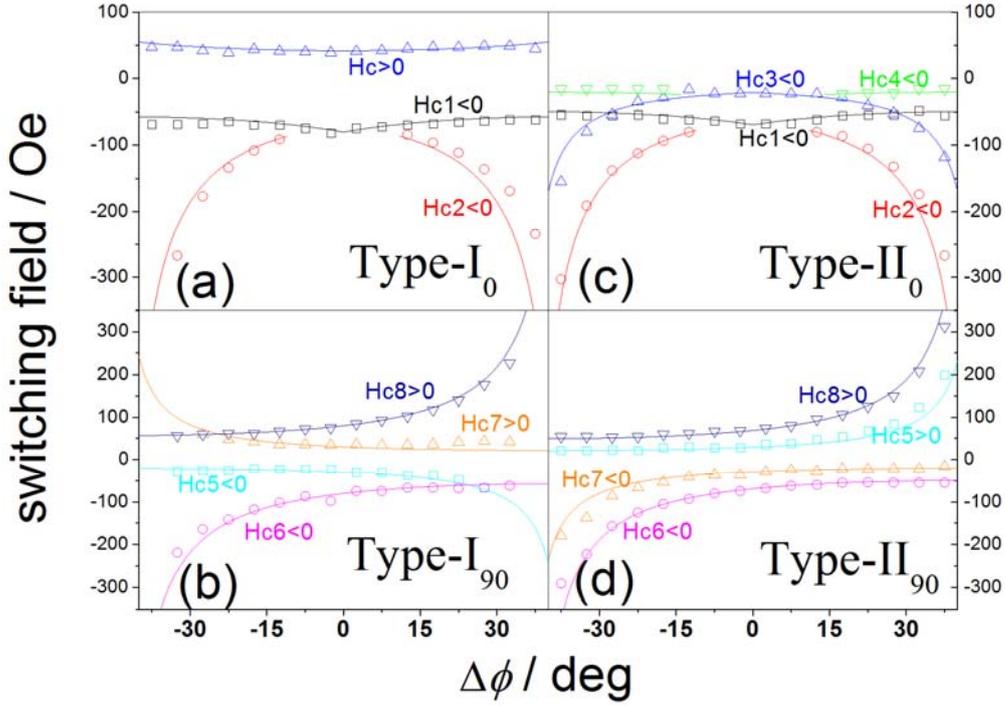

Fig. 3 (color online) Field orientation dependence of the experimentally observed switching fields (symbols) and the corresponding theoretical fittings (curves) for $A_{10}$ and $B_{10}$. $A_{10}$ exhibits the type-$I_0$ (a), and type-$I_{90}$ (b) angular dependent behaviors with the characterizing switching fields $H_{c1}<0$, $H_{c2}<0$, $H_c>0$ at $\phi_\parallel$, and $H_{c5}<0$, $H_{c6}<0$, $H_{c7}>0$, $H_{c8}>0$ at $\phi_\perp$. $B_{10}$ exhibits the type-$II_0$ (c), and type-$II_{90}$ (d) angular dependent behaviors with the characterizing switching fields $H_{c1}<0$, $H_{c2}<0$, $H_{c3}<0$, $H_{c4}<0$ at $\phi_\parallel$, and $H_{c5}>0$, $H_{c6}<0$, $H_{c7}<0$, $H_{c8}>0$ at $\phi_\perp$.

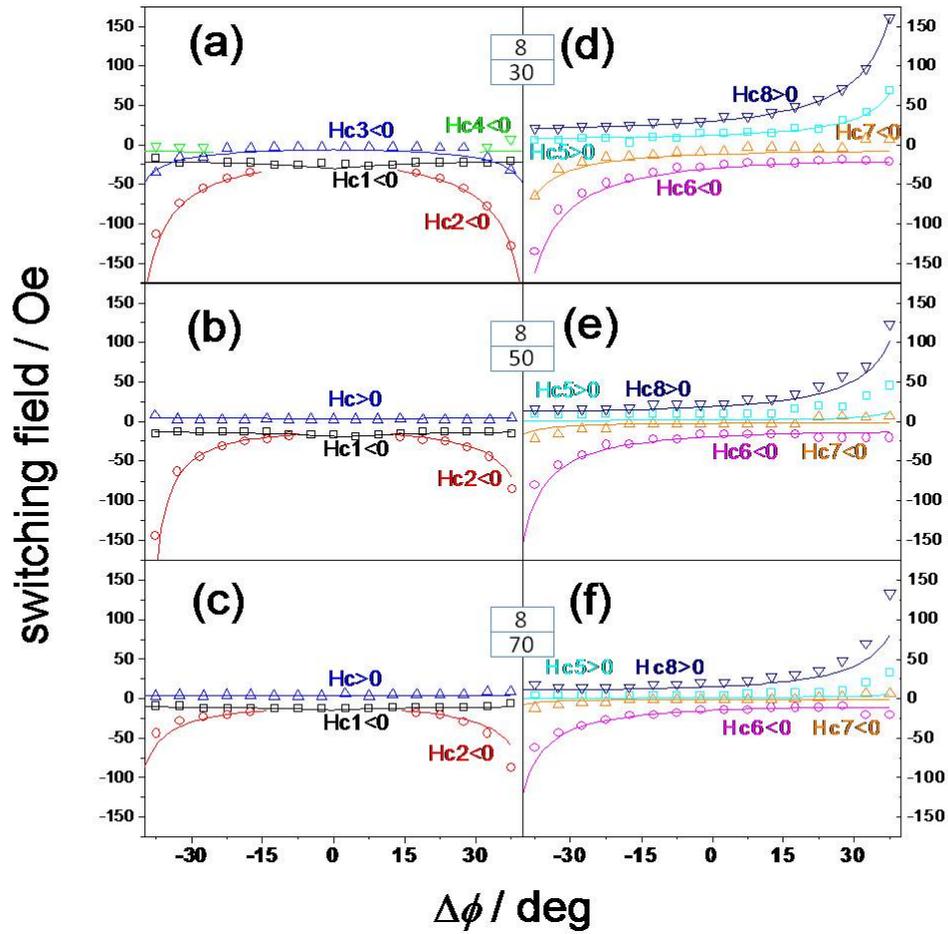

Fig. 4 (color online) Field orientation dependence of switching fields (symbols) and the corresponding theoretical fittings (curves) for $B_{30}$, $B_{50}$, and $B_{70}$ at $\phi_{\parallel}$ (a,b,c), and at $\phi_{\perp}$ (d,e,f), respectively. Transition from type-$II_0$ ($B_{30}$) to type-$I_0$ ($B_{50}$, $B_{70}$) was verified at $\phi_{\parallel}$; however, no corresponding transition was observed at $\phi_{\perp}$.

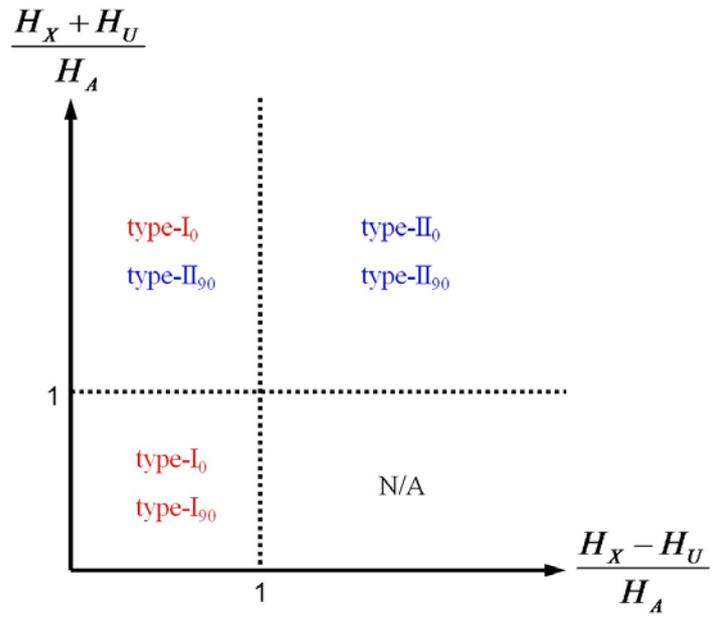

Fig. 5 (color online) Phase diagram showing the relationship between different types of angular dependent behaviors with the relative magnitudes of the effective anisotropy fields.